\documentclass[11pt]{article}
\usepackage{amssymb,amsfonts,amsmath,graphicx,hyperref}
\begin{document}

\title{Infinite--Dimensional Cerebellar Controller\\
for Realistic Human Biodynamics}
\author{Vladimir G. Ivancevic\thanks{%
Human Systems Integration, Land Operations Division, Defence Science \&
Technology Organisation, P.O. Box 1500, Edinburgh SA 5111, Australia
(Vladimir.Ivancevic@dsto.defence.gov.au)} \and Tijana T. Ivancevic\thanks{
School of Electrical and Information Engineering, University of South
Australia, Mawson Lakes, S.A. 5095, Australia (Tijana.Ivancevic@unisa.edu.au)%
}}
\date{}
\maketitle

\begin{abstract}
In this paper we propose an $\infty-$dimensional cerebellar model of neural controller for realistic human biodynamics. The model is developed using Feynman's action--amplitude (partition function) formalism. The cerebellum controller is acting as a supervisor for an autogenetic servo control of human musculo--skeletal dynamics, which is presented in (dissipative, driven) Hamiltonian form. The $\infty-$dimensional cerebellar controller is closely related to entropic motor control. \\

\noindent{\bf Keywords:} realistic human biodynamics, cerebellum motion control, $\infty-$dimensional neural network
\end{abstract}

\tableofcontents

\section{Introduction}

Realistic human biodynamics (RHB) is a science of human (and humanoid robot)
motion in its full complexity. It is governed by both Newtonian dynamics and
biological control laws (see \cite{Ganesh5,VladNick,GaneshSprSml,GaneshWSc,GaneshSprBig}).

There are over 200 bones in the human skeleton driven by about 640 muscular
actuators (see, e.g., \cite{Marieb}). While the muscles generate driving
torques in the moving joints,\footnote{%
Here we need to emphasize that human joints are significantly more flexible
than humanoid robot joints. Namely, each humanoid joint consists of a pair
of coupled segments with only Eulerian rotational degrees of freedom. On the
other hand, in each human synovial joint, besides gross Eulerian rotational
movements (roll, pitch and yaw), we also have some hidden and restricted
translations along $(X,Y,Z)-$axes. For example, in the knee joint, patella
(knee cap) moves for about 7--10 cm from maximal extension to maximal
flexion). It is well--known that even greater are translational amplitudes
in the shoulder joint. In other words, within the realm of rigid body
mechanics, a segment of a human arm or leg is not properly represented as a
rigid body fixed at a certain point, but rather as a rigid body hanging on
rope--like ligaments. More generally, the whole skeleton mechanically
represents a system of flexibly coupled rigid bodies. This implies the more
complex kinematics, dynamics and control then in the case of humanoid robots.%
} subcortical neural system performs both local and global (loco)motion
control: first reflexly controlling contractions of individual muscles, and
then orchestrating all the muscles into synergetic actions in order to
produce efficient movements. While the local reflex control of individual
muscles is performed on the \emph{spinal control level}, the global
integration of all the muscles into coordinated movements is performed
within the \emph{cerebellum} \cite{GaneshSprSml,GaneshWSc}.

All hierarchical subcortical neuro--muscular physiology, from the bottom
level of a single muscle fiber, to the top level of cerebellar muscular
synergy, acts as a \emph{temporal} $\langle {\rm out|in}\rangle$ \emph{reaction}, in such a way
that the higher level acts as a command/control space for the lower level,
itself representing an abstract image of the lower one:

\begin{enumerate}
\item At the \emph{muscular level}, we have \emph{excitation--contraction
dynamics} \cite{Hatze1,Hatze,Hatze2}, in which $\langle {\rm out|in}\rangle$ is given by the
following sequence of nonlinear diffusion processes \cite%
{GaneshSprSml,GaneshWSc}:
\begin{eqnarray*}
neural~action~potential~\rightsquigarrow synaptic~potential~\rightsquigarrow
muscular~action~potential \\
\rightsquigarrow~ excitation~contraction~coupling~\rightsquigarrow
muscle~tension~generating.
\end{eqnarray*}
Its purpose is the generation of muscular forces, to be transferred into
driving torques within the joint anatomical geometry.

\item At the \emph{spinal level}, $\langle {\rm out|in}\rangle$ is given by \emph{%
autogenetic--reflex stimulus--response control} \cite{Houk}. Here we have a
neural image of all individual muscles. The main purpose of the spinal
control level is to give both positive and negative feedbacks to stabilize
generated muscular forces within the `homeostatic' (or, more appropriately,
`homeokinetic') limits. The individual muscular actions are combined into
flexor--extensor (or agonist--antagonist) pairs, mutually controlling each
other. This is the mechanism of \emph{reciprocal innervation of agonists and
inhibition of antagonists}. It has a purely mechanical purpose to form the
so--called \emph{equivalent muscular actuators} (EMAs), which would generate
driving torques $T_{i}(t)$ for all movable joints.

\item At the \emph{cerebellar level}, $\langle {\rm out|in}\rangle$ is given by \emph{%
sensory--motor integration} \cite{HoukBarto}. Here we have an abstracted
image of all autogenetic reflexes. The main purpose of the cerebellar
control level is integration and fine tuning of the action of all active
EMAs into a synchronized movement, by \emph{supervising} the individual
autogenetic reflex circuits. At the same time, to be able to perform in new
and unknown conditions, the cerebellum is continuously adapting its own
neural circuitry by unsupervised (self--organizing) learning. Its action is
subconscious and automatic, both in humans and in animals.
\end{enumerate}

Naturally, we can ask the question: Can we assign a single $\langle {\rm out|in}\rangle$
measure to all these neuro--muscular stimulus--response reactions? We think
that we can do it; so in this Letter, we propose the concept of \emph{%
adaptive sensory--motor transition amplitude} as a unique measure for this
temporal $\langle {\rm out|in}\rangle$ relation. Conceptually, this $\langle {\rm out|in}\rangle-$\emph{amplitude}
can be formulated as the `\textit{neural path integral}':
\begin{equation}  \label{pijfj}
\langle {\rm out|in}\rangle\equiv\underset{\rm amplitude}{\left\langle {\rm motor|sensory}\right\rangle \,%
}=\int \mathcal{D}[w,x]\,\mathrm{e}^{\mathrm{i\,}S[x]}.
\end{equation}%
Here, the integral is taken over all \emph{activated} (or, `fired') \emph{%
neural pathways} $x^{i}=x^{i}(t)$ of the cerebellum, connecting its input $%
sensory-$state with its output $motor-$state, symbolically described by
\emph{adaptive neural measure} $\mathcal{D}[w,x]$, defined by the weighted
product (of discrete time steps)
\begin{equation}
\mathcal{D}[w,x]=\lim_{n\to\infty}\prod_{t=1}^{n}w_i(t)\,dx ^i(t), \label{prod1}
\end{equation}
in which the \emph{synaptic weights} $w_i=w_i(t)$, included in all active
neural pathways $x^{i}=x^{i}(t)$, are updated by the standard learning rule
\begin{equation*}
new\;value(t+1)\;=\;old\;value(t)\;+\;innovation(t).
\end{equation*}
More precisely, the weights $w_i$ in (\ref{prod1}) are updated according to one of the two
standard neural learning schemes, in which the micro--time level
is traversed in discrete steps, i.e., if $t=t_0,t_1,...,t_n$ then
$t+1=t_1,t_2,...,t_{n+1}$:\footnote{Note that we could also use a reward--based, {reinforcement
learning} rule \cite{SB}, in which system learns its {optimal
policy}:
$$
innovation(t)=|reward(t)-penalty(t)|.
$$}
\begin{enumerate}
    \item A \textit{self--organized}, \textit{unsupervised}
    (e.g., Hebbian--like \cite{Hebb}) learning rule:
\begin{equation}
w_i(t+1)=w_i(t)+ \frac{\sigma}{\eta}(w_i^{d}(t)-w_i^{a}(t)),
\label{Hebb}
\end{equation}
where $\sigma=\sigma(t),\,\eta=\eta(t)$ denote \textit{signal} and
\textit{noise}, respectively, while superscripts $d$ and $a$
denote \textit{desired} and \textit{achieved} micro--states,
respectively; or
    \item A certain form of a \textit{supervised gradient descent
    learning}:
\begin{equation}
w_i(t+1)\,=\,w_i(t)-\eta \nabla J(t), \label{gradient}
\end{equation}
where $\eta $ is a small constant, called the \textit{step size},
or the \textit{learning rate,} and $\nabla J(n)$ denotes the
gradient of the `performance hyper--surface' at the $t-$th
iteration.
\end{enumerate}
Theoretically, equations (\ref{pijfj}--\ref{gradient}) define an $\infty-$%
\emph{dimensional neural network} (see \cite{IA,IAY,NQJ}). Practically, in a computer simulation we
can use $10^7\leq n\leq 10^8$, roughly corresponding to the number of
neurons in the cerebellum \cite{NeuFuz,CompMind}.

The exponent term $S[x]$ in equation (\ref{pijfj}) represents the \emph{%
autogenetic--reflex action}, describing reflexly--induced motion of all
active EMAs, from their initial $stimulus-$state to their final $response-$%
state, along the family of extremal (i.e., Euler--Lagrangian) paths $x_{\min
}^{i}(t)$.\quad ($S[x]$ is properly derived in (\ref{affAct}--\ref{phaseint}%
) below.)

\section{Sub-Cerebellar Biodynamics and Its Spinal Reflex Servo--Control}

Subcerebellar biodynamics includes the following three components: (i) local
muscle--joint mechanics, (ii) whole--body musculo--skeletal dynamics, and
(iii) autogenetic reflex servo--control.

\subsection{Local Muscle--Joint Mechanics}

Local muscle--joint mechanics comprises of \cite%
{LieLagr,GaneshSprSml,GaneshWSc}):

{1. Synovial joint dynamics}, giving the first stabilizing effect to the
conservative skeleton dynamics, is described by the $(x,\dot{x})$--form of
the {Rayleigh -- Van der Pol's dissipation function}
\begin{equation*}
R=\frac{1}{2}\sum_{i=1}^{n}\,(\dot{x}^{i})^{2}\,[\alpha _{i}\,+\,\beta
_{i}(x^{i})^{2}],\quad
\end{equation*}
where $\alpha _{i}$ and $\beta _{i}$ denote dissipation parameters. Its
partial derivatives give rise to the viscous--damping torques and forces in
the joints
\begin{equation*}
\mathcal{F}_{i}^{joint}=\partial R/\partial \dot{x}^{i},
\end{equation*}
which are linear in $\dot{x}^{i}$ and quadratic in $x^{i}$.

{2. Muscular dynamics}, giving the driving torques and forces $\mathcal{F}%
_{i}^{muscle}=\mathcal{F}_{i}^{muscle}(t,x,\dot{ x})$ with $(i=1,\dots ,n)$
for RHB, describes the internal {excitation} and {contraction} dynamics of {%
equivalent muscular actuators} \cite{Hatze}.

(a) {Excitation dynamics} can be described by an impulse {force--time}
relation
\begin{eqnarray*}
F_{i}^{imp} &=&F_{i}^{0}(1\,-\,e^{-t/\tau _{i}})\text{ \qquad if stimulation
}>0 \\
\quad F_{i}^{imp} &=&F_{i}^{0}e^{-t/\tau _{i}}\qquad \qquad \quad\text{if
stimulation }=0,\quad
\end{eqnarray*}
where $F_{i}^{0}$ denote the maximal isometric muscular torques and forces,
while $\tau _{i}$ denote the associated time characteristics of particular
muscular actuators. This relation represents a solution of the Wilkie's
muscular {active--state element} equation \cite{Wilkie}
\begin{equation*}
\dot{\mu}\,+\,\gamma \,\mu \,=\,\gamma \,S\,A,\quad \mu (0)\,=\,0,\quad
0<S<1,
\end{equation*}
where $\mu =\mu (t)$ represents the active state of the muscle, $\gamma $
denotes the element gain, $A$ corresponds to the maximum tension the element
can develop, and $S=S(r)$ is the `desired' active state as a function of the
motor unit stimulus rate $r$. This is the basis for the RHB force controller.

(b) {Contraction dynamics} has classically been described by the Hill's {%
hyperbolic force--velocity }relation \cite{Hill}
\begin{equation*}
F_{i}^{Hill}\,=\,\frac{\left( F_{i}^{0}b_{i}\,-\,\delta _{ij}a_{i}\dot{x}%
^{j}\,\right) }{\left( \delta _{ij}\dot{x}^{j}\,+\,b_{i}\right) },\,\quad
\end{equation*}
where $a_{i}$ and $b_{i}$ denote the {Hill's parameters}, corresponding to
the energy dissipated during the contraction and the phosphagenic energy
conversion rate, respectively, while $\delta _{ij}$ is the Kronecker's $%
\delta-$tensor.

In this way, RHB describes the excitation/contraction dynamics for the $i$th
equivalent muscle--joint actuator, using the simple impulse--hyperbolic
product relation
\begin{equation*}
\mathcal{F}_{i}^{muscle}(t,x,\dot{x})=\,F_{i}^{imp}\times F_{i}^{Hill}.\quad
\end{equation*}

Now, for the purpose of biomedical engineering and rehabilitation, RHB has
developed the so--called {hybrid rotational actuator}. It includes, along
with muscular and viscous forces, the D.C. motor drives, as used in robotics
\cite{Vuk,LieLagr,GaneshSprSml}
\begin{equation*}
\mathcal{F}_{k}^{robo}=i_{k}(t)-J_{k}\ddot{x}_{k}(t)-B_{k}\dot{x}_{k}(t),
\end{equation*}
with
\begin{equation*}
l_{k}i_{k}(t)+R_{k}i_{k}(t)+C_{k}\dot{x}_{k}(t)=u_{k}(t),
\end{equation*}
where $k=1,\dots,n$, $i_{k}(t)$ and $u_{k}(t)$ denote currents and voltages
in the rotors of the drives, $R_{k},l_{k}$ and $C_{k}$ are resistances,
inductances and capacitances in the rotors, respectively, while $J_{k}$ and $%
B_{k}$ correspond to inertia moments and viscous dampings of the drives,
respectively.

Finally, to make the model more realistic, we need to add some stochastic
torques and forces \cite{GaneshIEEE,NeuFuz}
\begin{equation*}
\mathcal{F}_{i}^{stoch}=B_{ij}[x^{i}(t),t]\,dW^{j}(t)
\end{equation*}
where $B_{ij}[x(t),t]$ represents continuous stochastic {diffusion
fluctuations}, and $W^{j}(t)$ is an $N-$variable {Wiener process} (i.e.
generalized Brownian motion), with $dW^{j}(t)=W^{j}(t+dt)-W^{j}(t)$ for $%
j=1,\dots,N$.

\subsection{Hamiltonian Biodynamics and Its Reflex Servo--Control}

\begin{figure}[htb]
\centering \includegraphics[width=12cm]{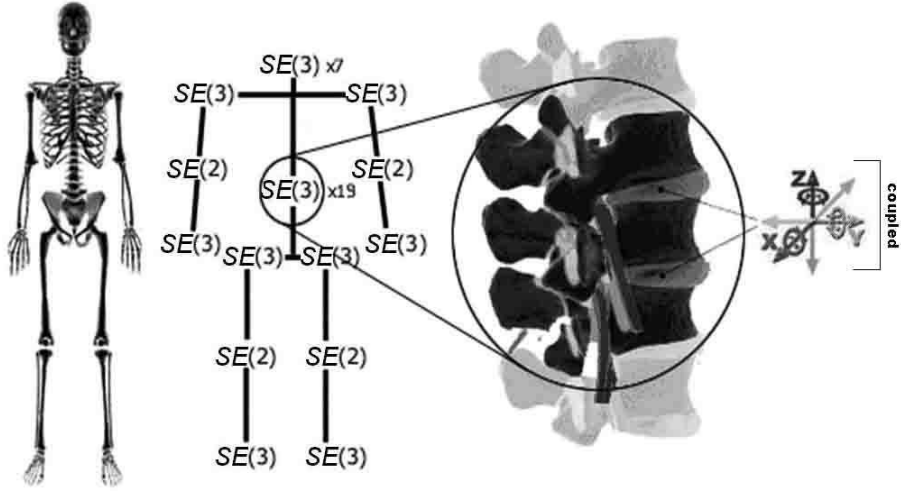}
\caption{The configuration manifold $Q$ of the human musculoskeletal
dynamics is defined as an anthropomorphic product of constrained Euclidean
SE(3)--groups acting in all major (synovial) human joints.}
\label{SpineSE(3)}
\end{figure}

General form of Hamiltonian biodynamics on the configuration manifold of
human motion is formulated in \cite%
{GaneshIEEE,IJMMS1,Ganesh5,VladNick,GaneshSprSml,GaneshSprBig}) using the
concept of Euclidean group of motions SE(3)\footnote{%
Briefly, the Euclidean SE(3)--group is defined as a semidirect
(noncommutative) product of 3D rotations and 3D translations, $%
SE(3):=SO(3)\rhd \mathbb{R}^{3}$. Its most important subgroups are the
following (for technical details see \cite{GaneshSprBig,ParkChung,GaneshADG}%
):
\par
{{\frame{$%
\begin{array}{cc}
\mathbf{Subgroup} & \mathbf{Definition} \\ \hline
\begin{array}{c}
SO(3),\text{ group of rotations} \\
\text{in 3D (a spherical joint)}%
\end{array}
&
\begin{array}{c}
\text{Set of all proper orthogonal } \\
3\times 3-\text{rotational matrices}%
\end{array}
\\ \hline
\begin{array}{c}
SE(2),\text{ special Euclidean group} \\
\text{in 2D (all planar motions)}%
\end{array}
&
\begin{array}{c}
\text{Set of all }3\times 3-\text{matrices:} \\
\left[
\begin{array}{ccc}
\cos \theta & \sin \theta & r_{x} \\
-\sin \theta & \cos \theta & r_{y} \\
0 & 0 & 1%
\end{array}%
\right]%
\end{array}
\\ \hline
\begin{array}{c}
SO(2),\text{ group of rotations in 2D} \\
\text{subgroup of }SE(2)\text{--group} \\
\text{(a revolute joint)}%
\end{array}
&
\begin{array}{c}
\text{Set of all proper orthogonal } \\
2\times 2-\text{rotational matrices} \\
\text{ included in }SE(2)-\text{group}%
\end{array}
\\ \hline
\begin{array}{c}
\mathbb{R}^{3},\text{ group of translations in 3D} \\
\text{(all spatial displacements)}%
\end{array}
& \text{Euclidean 3D vector space}%
\end{array}%
$}}}} (see Figure \ref{SpineSE(3)}),

Briefly, based on \emph{affine Hamiltonian function of human motion},
formally $H_{a}:T^{\ast }Q\rightarrow \mathbb{R},$ in local canonical
coordinates on the symplectic phase space (which is the cotangent bundle of
the human configuration manifold $Q$) $T^{\ast }Q$ given as
\begin{equation}
H_{a}(x,p,u)=H_{0}(x,p)-H^{j}(x,p)\,u_{j},  \label{aff}
\end{equation}%
where $H_{0}(x,p)=E_k(p)+E_p(x)$ is the physical Hamiltonian (kinetic +
potential energy) dependent on joint coordinates $x^{i}$ and their canonical
momenta $p_{i}$, $H^{j}=H^{j}(x,p)$, ($j=1,\dots ,\,m\leq n$ are the
coupling Hamiltonians corresponding to the system's active joints and $%
u_{i}=u_{i}(t,x,p)$ are (reflex) feedback--controls. Using (\ref{aff}) we
come to the affine Hamiltonian control RHB--system, in deterministic form%
\begin{align}
\dot{x}^{i}& =\partial _{p_{i}}H_{0}-\partial _{p_{i}}H^{j}\,u_{j}+\partial
_{p_{i}}R,  \label{af1} \\
\dot{p}_{i}& =\mathcal{F}_{i}-\partial _{x^{i}}H_{0}+\partial
_{x^{i}}H^{j}\,u_{j}+\partial _{x^{i}}R,  \notag \\
o^{i}& =-\partial _{u_{i}}H_{a}=H^{j},  \notag \\
x^{i}(0)& =x_{0}^{i},\qquad p_{i}(0)=p_{i}^{0},  \notag \\
(i& =1,\dots ,n;\qquad j=1,\dots ,\,Q\leq n),  \notag
\end{align}%
(where $\partial _{u}\equiv \partial /\partial u$, $\mathcal{F}_{i}=\mathcal{%
F}_{i}(t,x,p),$ $H_{0}=H_{0}(x,p),$ $H^{j}=H^{j}(x,p),$ $H_{a}=H_{a}(x,p,u),$
$R=R(x,p)$), as well as in the fuzzy--stochastic form \cite%
{GaneshIEEE,NeuFuz}%
\begin{align}
dq^{i}& =\left( \partial _{p_{i}}H_{0}(\sigma _{\mu })-\partial
_{p_{i}}H^{j}(\sigma _{\mu })\,u_{j}+\partial _{p_{i}}R\right) \,dt,  \notag
\\
dp_{i}& =B_{ij}[x^{i}(t),t]\,dW^{j}(t)\qquad +\qquad \qquad  \label{af2} \\
& \left( \bar{\mathcal{F}}_{i}-\partial _{x^{i}}H_{0}(\sigma _{\mu
})+\partial _{x^{i}}H^{j}(\sigma _{\mu })\,u_{j}+\partial _{x^{i}}R\right)
\,dt,  \notag \\
d\bar{o}^{i}& =-\partial _{u_{i}}H_{a}(\sigma _{\mu })\,dt=H^{j}(\sigma
_{\mu })\,dt,\qquad \qquad  \notag \\
x^{i}(0)& =\bar{x}_{0}^{i},\qquad p_{i}(0)=\bar{p}_{i}^{0}\qquad \qquad
\notag
\end{align}%
In (\ref{af1})--(\ref{af2}), $R=R(x,p)$ denotes the joint (nonlinear)
dissipation function, $o^{i}$ are affine system outputs (which can be
different from joint coordinates); $\{\sigma \}_{\mu }$ \ (with $\mu \geq 1$%
) denote fuzzy sets of conservative parameters (segment lengths, masses and
moments of inertia), dissipative joint dampings and actuator parameters
(amplitudes and frequencies), while the bar $\bar{(.)}$ over a variable
denotes the corresponding fuzzified variable; $B_{ij}[q^{i}(t),t]$ denote
diffusion fluctuations and $W^{j}(t)$ are discontinuous jumps as the $n$%
--dimensional Wiener process.

In this way, the {force RHB servo--controller} is formulated as affine
control Hamiltonian--systems (\ref{af1}--\ref{af2}), which resemble the \emph{autogenetic motor servo} (see Appendix), acting on the spinal--reflex level of
the human locomotion control. A voluntary contraction force $F$ of human
skeletal muscle is reflexly excited (positive feedback $+F^{-1}$) by the
responses of its {spindle receptors} to stretch and is reflexly inhibited
(negative feedback $-F^{-1}$) by the responses of its {Golgi tendon organs}
to contraction. Stretch and unloading reflexes are mediated by combined
actions of several autogenetic neural pathways, forming the so--called $%
\mathbf{`}$motor servo.' The term $\mathbf{`}$autogenetic' means that the
stimulus excites receptors located in the same muscle that is the target of
the reflex response. The most important of these muscle receptors are the
primary and secondary endings in the muscle--spindles, which are sensitive
to length change -- positive length feedback $+F^{-1}$, and the Golgi tendon
organs, which are sensitive to contractile force -- negative force feedback $%
-F^{-1}$.

The gain $G$ of the length feedback $+F^{-1}$ can be expressed as the {%
positional stiffness} (the ratio $G\approx S=dF/dx$ of the force--$F$ change
to the length--$x$ change) of the muscle system. The greater the stiffness $S
$, the less the muscle will be disturbed by a change in load. The
autogenetic circuits $+F^{-1}$ and $-F^{-1}$ appear to function as {%
servoregulatory loops} that convey continuously graded amounts of excitation
and inhibition to the large ({alpha}) skeletomotor neurons. Small ({gamma})
fusimotor neurons innervate the contractile poles of muscle spindles and
function to modulate spindle--receptor discharge.

\section{Cerebellum: The Adaptive Path--Integral Comparator}

\subsection{Cerebellum as a Neural Controller}

\begin{figure}[ht]
\centerline{\includegraphics[height=8cm]{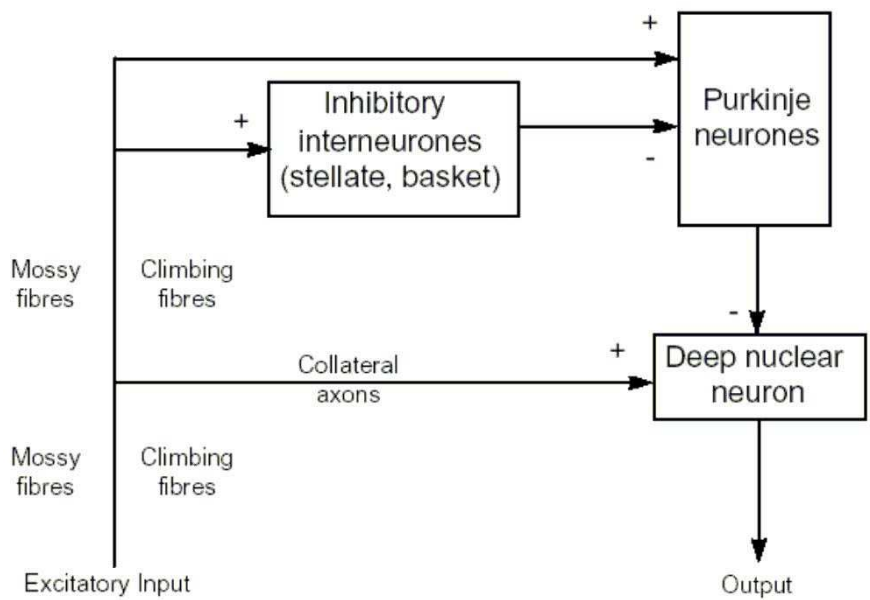}}
\caption{Schematic $\langle {\rm out|in}\rangle$ organization of the primary cerebellar
circuit. In essence, excitatory inputs, conveyed by collateral axons of
Mossy and Climbing fibers activate directly neurones in the Deep cerebellar
nuclei. The activity of these latter is also modulated by the inhibitory
action of the cerebellar cortex, mediated by the Purkinje cells.}
\label{cerebellum1}
\end{figure}

Having, thus, defined the spinal reflex control level, we proceed to model
the top subcortical commander/controller, the \emph{cerebellum} (see Appendix).
The cerebellum is responsible for coordinating precisely timed $\langle {\rm out|in}\rangle$
activity by integrating motor output with ongoing sensory feedback (see Figure \ref{cerebellum1}). It
receives extensive projections from sensory--motor areas of the cortex and
the periphery and directs it back to premotor and motor cortex \cite%
{Ghez1,Ghez2}. This suggests a role in sensory--motor integration and the
timing and execution of human movements. The cerebellum stores patterns of
motor control for frequently performed movements, and therefore, its
circuits are changed by experience and training. It was termed the \emph{%
adjustable pattern generator} in the work of J. Houk and collaborators \cite%
{HoukBarto}. Also, it has become the inspiring `brain--model' in
robotic research \cite{sch1,sch3,sch2}.

The cerebellum is known to be involved in the
production and learning of smooth coordinated movements \cite{Thach,Faag}. Two classes of inputs carry information into the
cerebellum: the mossy fibers (MFs) and the climbing
fibers (CFs). The MFs provide both plant state
and contextual information \cite{Bloedel}. The CFs, on the other
hand, are thought to provide information that reflect
errors in recently generated movements \cite{Ito84,Ito90}. This information
is used to adjust the programs encoded by
the cerebellum. The MFs carry plant state, motor efference, and other contextual signals
into the cerebellum. These fibers impinge on
granule cells, whose axons give rise to parallel fibers
(PFs). Through the combination of inputs from multiple
classes of MFs and local inhibitory interneurons,
the granule cells are thought to provide a sparse expansive encoding of the incoming state information \cite{Albus}. The large number of PFs converge on a
much smaller set of Purkinje cells (PCs), while the PCs, in turn, provide inhibitory
signals to a single cerebellar nuclear cell \cite{Faag}.
Using this principle, the Cerebellar Model Arithmetic Computer, or CMAC--neural network has been built \cite{Albus,Miller} and implemented in robotics \cite{Smagt}, using trial-and-error learning to produce bursts of muscular activity for controlling
robot arms.

So, this `cerebellar control' works for simple robotic problems, like non-redundant manipulation. However, comparing the number of its neurons ($10^7-10^8$), to the size of
conventional neural networks (including CMAC), suggests that artificial neural nets \emph{cannot} satisfactorily model the function of this sophisticated
`super--bio--computer', as its dimensionality is virtually infinite. Despite
a lot of research dedicated to its structure and function (see \cite{HoukBarto} and references there cited), the real nature of the cerebellum
still remains a `mystery'.
\begin{figure}[ht]
\centerline{\includegraphics[width=15cm]{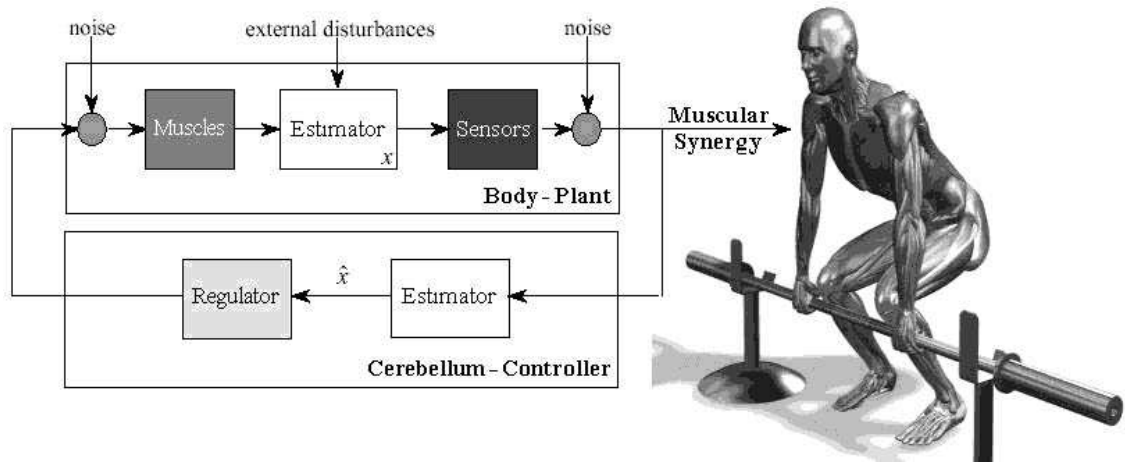}}
\caption{The cerebellum as a motor controller.}
\label{CrblCtrl}
\end{figure}

The main function of the cerebellum as a motor controller is depicted in
Figure \ref{CrblCtrl}. A coordinated movement is easy to recognize, but we
know little about how it is achieved. In search of the neural basis of
coordination, a model of spinocerebellar interactions was recently presented
in \cite{Apps}, in which the structural and functional organizing principle
is a division of the cerebellum into discrete micro--complexes. Each
micro--complex is the recipient of a specific motor error signal, that is, a
signal that conveys information about an inappropriate movement. These
signals are encoded by spinal reflex circuits and conveyed to the cerebellar
cortex through climbing fibre afferents. This organization reveals salient
features of cerebellar information processing, but also highlights the
importance of systems level analysis for a fuller understanding of the
neural mechanisms that underlie behavior.

\subsection{Hamiltonian Action and Neural Path Integral}

Here, we propose a \emph{quantum--like adaptive control} approach to
modeling the `cerebellar mystery'. Corresponding to the affine Hamiltonian
control function (\ref{aff}) we define the \emph{affine Hamiltonian control
action},
\begin{equation}
S_{aff}[q,p]=\int_{t_{in}}^{t_{out}}d\tau \left[ p_{i}\dot{q}%
^{i}-H_{aff}(q,p)\right].  \label{affAct}
\end{equation}

From the affine Hamiltonian action (\ref{affAct}) we further derive the
associated expression for the \emph{neural phase--space path integral} (in
normal units), representing the \emph{cerebellar sensory--motor amplitude} $%
\langle {\rm out|in}\rangle$,
\begin{eqnarray}
\left\langle q_{out}^{i},p_{i}^{out}|q_{in}^{i},p_{i}^{in}\right\rangle
&=&\int \mathcal{D}[w,q,p]\,\mathrm{e}^{\mathrm{i\,}S_{aff}[q,p]}\qquad
\qquad \qquad  \label{phaseint} \\
&=&\int \mathcal{D}[w,q,p]\exp \left\{ \mathrm{i}\int_{t_{in}}^{t_{out}}d%
\tau \left[ p_{i}\dot{q}^{i}-H_{aff}(q,p)\right] \right\} ,  \notag \\
\qquad\text{with}\qquad &&\int \mathcal{D}[w,q,p]=\int \prod_{\tau =1}^{n}%
\frac{w^{i}(\tau )dp_{i}(\tau )dq^{i}(\tau )}{2\pi },  \notag
\end{eqnarray}
where $w_i=w_i(t)$ denote the cerebellar synaptic weights positioned along
its neural pathways, being continuously updated using the Hebbian--like
self--organizing learning rule (\ref{Hebb}). Given the transition
amplitude $out|in$ (\ref{phaseint}), the \emph{cerebellar sensory--motor
transition probability} is defined as its absolute square, $|\langle {\rm out|in}\rangle|^2$.

In the phase--space path integral (\ref{phaseint}), $q_{in}^{i}=q_{in}^{i}(t),\;
q_{out}^{i}=q_{out}^{i}(t);\; p^{in}_{i}=p^{in}_{i}(t),\;
p^{out}_{i}=p^{out}_{i}(t);\; t_{in}\leq t\leq t_{out}$, for all discrete
time steps, $t=1,...,n\to\infty$, and we are allowing for the affine
Hamiltonian $H_{aff}(q,p)$ to depend upon all the ($M\leq N$) EMA--angles
and angular momenta collectively. Here, we actually systematically took a
discretized differential time limit of the form $t_{\sigma }-t_{\sigma
-1}\equiv d\tau $ (both $\sigma $ and $\tau $ denote discrete time steps)
and wrote $\frac{(q_{\sigma }^{i}-q_{\sigma -1}^{i})}{(t_{\sigma }-t_{\sigma
-1})}\equiv \dot{q}^{i}$. For technical details regarding the path integral
calculations on Riemannian and symplectic manifolds (including the standard
regularization procedures), see \cite{Klauder,Klauder1}.

Now, motor learning occurring in the cerebellum can be observed using
functional MR imaging, showing changes in the cerebellar action potential,
related to the motor tasks (see, e.g., \cite{Mascalchi}). To account for
these electro--physiological currents, we need to add the \emph{source} term
$J_i(t)q^i(t)$ to the affine Hamiltonian action (\ref{affAct}), (the current
$J_i=J_i(t)$ acts as a source $J_iA^i$ of the \emph{cerebellar electrical
potential} $A^i=A^i(t)$),
\begin{equation*}
S_{aff}[q,p,J]=\int_{t_{in}}^{t_{out}}d\tau \left[ p_{i}\dot{q}%
^{i}-H_{aff}(q,p)+J_iq^i\right],
\end{equation*}
which, subsequently gives the cerebellar path integral with the action
potential source, coming either from the motor cortex or from other
subcortical areas.

Note that the standard \emph{Wick rotation}: $t\mapsto{} t$ (see \cite%
{Klauder,Klauder1}), makes our path integral real, i.e.,
\begin{equation*}
\int \mathcal{D}[w,q,p]\,\mathrm{e}^{\mathrm{i\,}S_{aff}[q,p]}\quad
\underrightarrow{Wick}\quad \int \mathcal{D}[w,q,p]\,\mathrm{e}^{\mathrm{-\,}%
S_{aff}[q,p]},
\end{equation*}
while their subsequent discretization gives the standard thermodynamic \emph{%
partition function} (see Appendix),
\begin{equation}
Z=\sum_j{}^{-w_jE^j/T},  \label{partition}
\end{equation}
where $E^j$ is the energy eigenvalue corresponding to the affine Hamiltonian
$H_{aff}(q,p)$, $T$ is the temperature--like environmental control
parameter, and the sum runs over all energy eigenstates (labelled by the
index $j$). From (\ref{partition}), we can further calculate all statistical
and thermodynamic system properties (see \cite{FeynmanStat}), as for
example, \emph{transition entropy} $S = k_B\ln Z$, etc.

\subsection{Entropy and Motor Control}

Our cerebellar path integral controller is closely related to \textit{entropic motor control} \cite{Hong1,Hong2}, which deals with
neuro-physiological feedback information and environmental uncertainty. The
probabilistic nature of human motor action can be characterized by entropies
at the level of the organism, task, and environment. Systematic changes in
motor adaptation are characterized as task--organism and
environment--organism tradeoffs in entropy. Such compensatory adaptations
lead to a view of goal--directed motor control as the product of an
underlying conservation of entropy across the task--organism--environment
system. In particular, an experiment conducted in \cite{Hong2} examined the
changes in entropy of the coordination of isometric force output under
different levels of task demands and feedback from the environment. The goal
of the study was to examine the hypothesis that human motor adaptation can
be characterized as a process of entropy conservation that is reflected in
the compensation of entropy between the task, organism motor output, and
environment. Information entropy of the coordination dynamics relative phase
of the motor output was made conditional on the idealized situation of human
movement, for which the goal was always achieved. Conditional entropy of the
motor output decreased as the error tolerance and feedback frequency were
decreased. Thus, as the likelihood of meeting the task demands was decreased
increased task entropy and/or the amount of information from the environment
is reduced increased environmental entropy, the subjects of this experiment
employed fewer coordination patterns in the force output to achieve the
goal. The conservation of entropy supports the view that context dependent
adaptations in human goal--directed action are guided fundamentally by
natural law and provides a novel means of examining human motor behavior.
This is fundamentally related to the \textit{Heisenberg uncertainty principle%
} \cite{QuLeap} and further supports the argument for the primacy of a
probabilistic approach toward the study of biodynamic cognition systems.

The action--amplitude formalism represents a kind of a generalization of the
Haken-Kelso-Bunz (HKB) model of self-organization in the individual's motor
system \cite{HKB,Kelso95}, including: multi-stability, phase
transitions and hysteresis effects, presenting a contrary view to
the purely feedback driven systems. HKB uses the concepts of
synergetics (order parameters, control parameters, instability,
etc) and the mathematical tools of nonlinearly coupled (nonlinear)
dynamical systems to account for self-organized behavior both at
the cooperative, coordinative level and at the level of the
individual coordinating elements. The HKB model stands as a
building block upon which numerous extensions and elaborations
have been constructed. In particular, it has been possible to
derive it from a realistic model of the cortical sheet in which
neural areas undergo a reorganization that is mediated by intra-
and inter-cortical connections. Also, the HKB model describes
phase transitions (`switches') in coordinated human movement as
follows: (i) when the agent begins in the anti-phase mode and
speed of movement is increased, a spontaneous switch to
symmetrical, in-phase movement occurs; (ii) this transition
happens swiftly at a certain critical frequency; (iii) after the
switch has occurred and the movement rate is now decreased the
subject remains in the symmetrical mode, i.e. she does not switch
back; and (iv) no such transitions occur if the subject begins
with symmetrical, in-phase movements. The HKB dynamics of the
order parameter relative phase as is given by a nonlinear
first-order ODE:
$$\dot{\phi} = (\alpha + 2 \beta r^2) \sin\phi - \beta r^2
\sin2\phi,
$$
where $\phi$ is the phase relation (that characterizes the
observed patterns of behavior, changes abruptly at the transition
and is only weakly dependent on parameters outside the phase
transition), $r$ is the oscillator amplitude, while $\alpha,\beta$
are coupling parameters (from which the critical frequency where
the phase transition occurs can be calculated).

\section{Appendix}

\subsection{Houk's Autogenetic Motor Servo}

\begin{figure}[h]
\centering
\includegraphics[width=13cm]{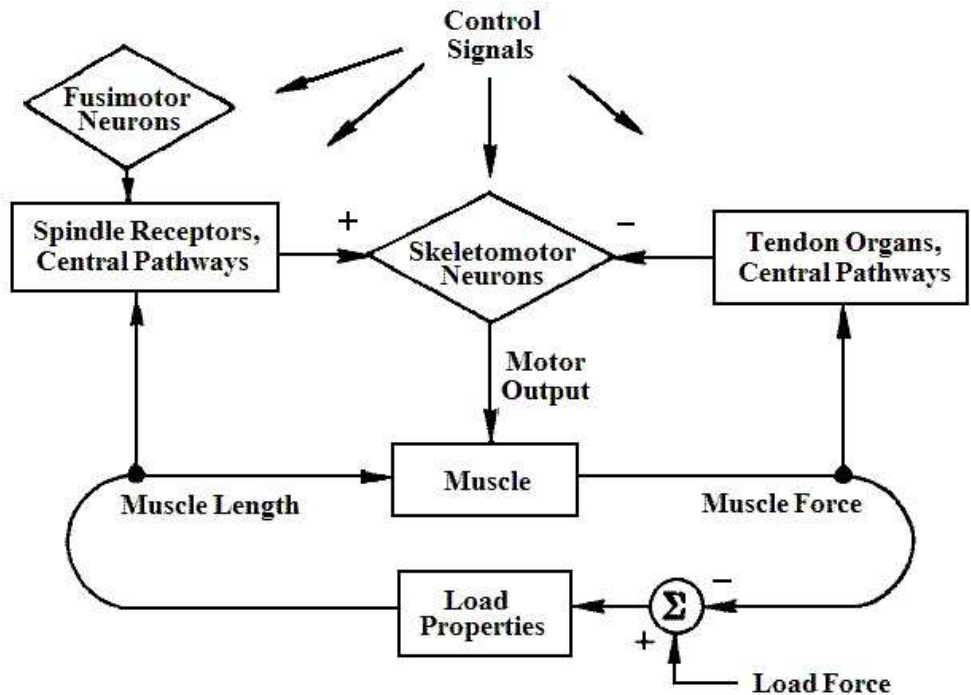}\caption{Houk's
autogenetic motor servo.}\label{HoukFigure}
\end{figure}

About three decades ago, James Houk pointed out in
\cite{Houk67,Houk70,Houk78,Houk79} that stretch and unloading
reflexes were mediated by combined actions of several autogenetic
neural pathways. In this context, ``autogenetic" (or, autogenic)
means that the stimulus excites receptors located in the same
muscle that is the target of the reflex response. The most
important of these muscle receptors are the primary and secondary
endings in muscle spindles, sensitive to length change, and the
Golgi tendon organs, sensitive to contractile force. The
autogenetic circuits appear to function as servo-regulatory loops
that convey continuously graded amounts of excitation and
inhibition to the large (alpha) skeletomotor neurons. Small
(gamma) fusimotor neurons innervate the contractile poles of
muscle spindles and function to modulate spindle--receptor
discharge. Houk's term ``motor servo" \cite{Houk78} has been used
to refer to this entire control system, summarized by the block
diagram in Figure \ref{HoukFigure}.

Prior to a study by Matthews \cite{Matthews69}, it was widely
assumed that secondary endings belong to the mixed population of
``flexor reflex afferents," so called because their activation
provokes the flexor reflex pattern -- excitation of flexor
motoneurons and inhibition of extensor motoneurons. Matthews'
results indicated that some category of muscle stretch receptor
other than the primary ending provides important excitation to
extensor muscles, and he argued forcefully that it must be the
secondary ending.

The primary and secondary muscle spindle afferent fibers both
arise from a specialized structure within the muscle, the
\emph{muscle spindle}, a fusiform structure 4--7 mm long and
80--200 $\mu$ in diameter. The spindles are located deep within
the muscle mass, scattered widely through the muscle body, and
attached to the tendon, the endomysium or the perimysium, so as to
be in parallel with the extrafusal or regular muscle fibers.
Although spindles are scattered widely in muscles, they are not
found throughout. Muscle spindle (see Figure \ref{mSpindle})
contains two types of intrafusal muscle fibers (intrafusal means
inside the fusiform spindle): the nuclear bag fibers and the
nuclear chain fibers. The nuclear bag fibers are thicker and
longer than the nuclear chain fibers, and they receive their name
from the accumulation of their nuclei in the expanded bag-like
equatorial region-the nuclear bag. The nuclear chain fibers have
no equatorial bulge; rather their nuclei are lined up in the
equatorial region-the nuclear chain. A typical spindle contains
two nuclear bag fibers and 4-5 nuclear chain fibers.

The pathways from primary and secondary endings are treated
commonly by Houk in Figure \ref{HoukFigure}, since both receptors
are sensitive to muscle length and both provoke reflex excitation.
However, primary endings show an additional sensitivity to the
dynamic phase of length change, called dynamic responsiveness, and
they also show a much--enhanced sensitivity to small changes in
muscle length \cite{Matthews72}.

The motor servo comprises three closed circuits (Figure
\ref{HoukFigure}), two neural feedback pathways, and one circuit
representing the mechanical interaction between a muscle and its
load. One of the feedback pathways, that from spindle receptors,
conveys information concerning muscle length, and it follows that
this loop will act to keep muscle length constant. The other
feedback pathway, that from tendon organs, conveys information
concerning muscle force, and it acts to keep force constant.

In general, it is physically impossible to maintain both muscle
length and force constant when external loads vary; in this
situation the action of the two feedback loops will oppose each
other. For example, an increased load force will lengthen the
muscle and cause muscular force to increase as the muscle is
stretched out on its length-tension curve. The increased length
will lead to excitation of motoneurons, whereas the increased
force will lead to inhibition. It follows that the net regulatory
action conveyed by skeletomotor output will depend on some
relationship between force change and length change and on the
strength of the feedback from muscle spindles and tendon organs. A
simple mathematical derivation \cite{Nichols76} demonstrates that
the change in skeletomotor output, the error signal of the motor
servo, Should be proportional to the difference between a
regulated stiffness and the actual stiffness provided by the
mechanical properties of the muscle, where stiffness has the units
of force change divided by length change. The regulated stiffness
is determined by the ratio of the gain of length to force
feedback.

It follows that the combination of spindle receptor and tendon
organ feedback will tend to maintain the stiffness of the
neuromuscular apparatus at some regulated level. If this level is
high, due to a high gain of length feedback and a low gain of
force feedback, one could simply forget about force feedback and
treat muscle length as the regulated variable of the system.
However, if the regulated level of stiffness is intermediate in
value, i.e. not appreciably different from the average stiffness
arising from muscle mechanical properties in the absence of reflex
actions, one would conclude that stiffness, or its inverse,
compliance, is the regulated property of the motor servo.

In this way, the autogenetic reflex motor servo provides the
local, reflex feedback loops for individual muscular contractions.
A voluntary contraction force $F$ of human skeletal muscle is
reflexly excited (positive feedback $ +F^{-1}$) by the responses
of its \textit{spindle receptors} to stretch and is reflexly
inhibited (negative feedback $-F^{-1}$) by the responses of its
\textit{Golgi tendon organs} to contraction. Stretch and unloading
reflexes are mediated by combined actions of several autogenetic
neural pathways, forming the \textit{motor servo} (see
\cite{GaneshSprSml,GaneshWSc,GaneshSprBig}).

In other words, branches of the afferent fibers also synapse with
with interneurons that inhibit motor neurons controlling the
antagonistic muscles -- \textit{reciprocal inhibition}.
Consequently, the stretch stimulus causes the antagonists to relax
so that they cannot resists the shortening of the stretched muscle
caused by the main reflex arc. Similarly, firing of the Golgi
tendon receptors causes inhibition of the muscle contracting too
strong and simultaneous \textit{reciprocal activation} of its
antagonist.

\subsection{Cerebellum and Muscular Synergy}

The cerebellum is a
brain region anatomically located at the bottom rear of the head (the
hindbrain), directly above the brainstem, which is important for a number of
subconscious and automatic motor functions, including motor learning. It
processes information received from the motor cortex, as well as from
proprioceptors and visual and equilibrium pathways, and gives `instructions'
to the motor cortex and other subcortical motor centers (like the basal
nuclei), which result in proper balance and posture, as well as smooth,
coordinated skeletal movements, like walking, running, jumping, driving,
typing, playing the piano, etc. Patients with cerebellar dysfunction have
problems with precise movements, such as walking and balance, and hand and
arm movements. The cerebellum looks \emph{similar in all animals}, from fish
to mice to humans. This has been taken as evidence that it performs a common
function, such as regulating motor learning and the timing of movements, in
all animals. Studies of simple forms of motor learning in the
vestibulo--ocular reflex and eye--blink conditioning are demonstrating that
timing and amplitude of learned movements are encoded by the cerebellum.

When someone compares learning a new skill to learning how to ride
a bike they imply that once mastered, the task seems imbedded in
our brain forever. Well, imbedded in the cerebellum to be exact.
This brain structure is the commander of coordinated movement and
possibly even some forms of cognitive learning. Damage to this
area leads to motor or movement difficulties.

A part of a human brain that is devoted to the sensory-motor
control of human movement, that is motor coordination and
learning, as well as equilibrium and posture, is the cerebellum
(which in Latin means ``little brain"). It performs integration of
sensory perception and motor output. Many neural pathways link the
cerebellum with the motor cortex, which sends information to the
muscles causing them to move, and the spino--cerebellar tract,
which provides proprioception, or feedback on the position of the
body in space. The cerebellum integrates these pathways, using the
constant feedback on body position to fine--tune motor movements
\cite{Ito84}.

The human cerebellum has 7--14 million Purkinje cells. Each
receives about 200,000 synapses, most onto dendritic splines.
Granule cell axons form the \emph{parallel fibers}. They make
excitatory synapses onto Purkinje cell dendrites. Each parallel
fibre synapses on about 200 Purkinje cells. They create a strip of
excitation along the cerebellar folia.

\begin{figure}[h]
\centerline{\includegraphics[width=10cm]{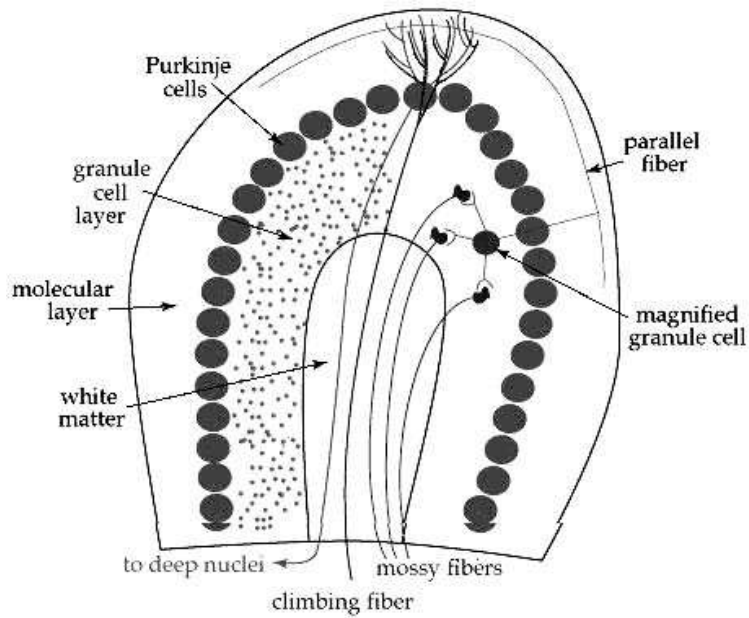}}
\caption{Stereotypical ways throughout the cerebellum.}
\label{cb9}
\end{figure}

\emph{Mossy fibers} are one of two main sources of input to the
cerebellar cortex (see Figure \ref{cb9}). A mossy fibre is an axon
terminal that ends in a large, bulbous swelling. These mossy
fibers enter the granule cell layer and synapse on the dendrites
of granule cells; in fact the granule cells reach out with little
`claws' to grasp the terminals. The granule cells then send their
axons up to the molecular layer, where they end in a T and run
parallel to the surface. For this reason these axons are called
\emph{parallel fibers}. The parallel fibers synapse on the huge
dendritic arrays of the Purkinje cells. However, the individual
parallel fibers are not a strong drive to the Purkinje cells. The
Purkinje cell dendrites fan out within a plane, like the splayed
fingers of one hand. If we were to turn a Purkinje cell to the
side, it would have almost no width at all. The parallel fibers
run perpendicular to the Purkinje cells, so that they only make
contact once as they pass through the dendrites.

Unless firing in bursts, parallel fibre EPSPs do not fire Purkinje
cells. Parallel fibers provide excitation to all of the Purkinje
cells they encounter. Thus, granule cell activity results in a
strip of activated Purkinje cells.

Mossy fibers arise from the spinal cord and brainstem. They
synapse onto granule cells and deep cerebellar nuclei. The
Purkinje cell makes an inhibitory synapse (GABA) to the deep
nuclei. Mossy fibre input goes to both cerebellar cortex and deep
nuclei. When the Purkinje cell fires, it inhibits output from the
deep nuclei.

The \textit{climbing fibre} arises from the inferior olive. It
makes about 300 excitatory synapses onto one Purkinje cell. This
powerful input can fire the Purkinje cell.

The parallel fibre synapses are plastic---that is, they can be
modified by experience. When parallel fibre activity and climbing
fibre activity converge on the same Purkinje cell, the parallel
fibre synapses become weaker (EPSPs are smaller). This is called
long-term depression. Weakened parallel fibre synapses result in
less Purkinje cell activity and less inhibition to the deep
nuclei, resulting in facilitated deep nuclei output. Consequently,
the mossy fibre collaterals control the deep nuclei.

The \textit{basket cell} is activated by parallel fibers
afferents. It makes inhibitory synapses onto Purkinje cells. It
provides lateral inhibition to Purkinje cells. Basket cells
inhibit Purkinje cells lateral to the active beam.

\textit{Golgi cells} receive input from parallel fibers, mossy
fibers, and climbing fibers. They inhibit granule cells. Golgi
cells provide feedback inhibition to granule cells as well as
feedforward inhibition to granule cells. Golgi cells create a
brief burst of granule cell activity.

Although each parallel fibre touches each Purkinje cell only once,
the thousands of parallel fibers working together can drive the
Purkinje cells to fire like mad.

The second main type of input to the folium is the \emph{climbing
fibre}. The climbing fibers go straight to the Purkinje cell layer
and snake up the Purkinje dendrites, like ivy climbing a trellis.
Each climbing fibre associates with only one Purkinje cell, but
when the climbing fibre fires, it provokes a large response in the
Purkinje cell.

The Purkinje cell compares and processes the varying inputs it
gets, and finally sends its own axons out through the white matter
and down to the \textit{deep nuclei}. Although the inhibitory
Purkinje cells are the main output of the cerebellar cortex, the
output from the cerebellum as a whole comes from the deep nuclei.
The three deep nuclei are responsible for sending excitatory
output back to the thalamus, as well as to postural and vestibular
centers.

There are a few other cell types in cerebellar cortex, which can
all be lumped into the category of inhibitory interneuron. The
\emph{Golgi cell} is found among the granule cells. The
\emph{stellate} and \emph{basket cells} live in the molecular
layer. The basket cell (right) drops axon branches down into the
Purkinje cell layer where the branches wrap around the cell bodies
like baskets.

The cerebellum operates in 3's: there are 3 highways leading in
and out of the cerebellum, there are 3 main inputs, and there are
3 main outputs from 3 deep nuclei. They are:

The 3 highways are the \textit{peduncles}. There are 3 pairs (see
\cite{Molavi,Harting,Marieb}):
\begin{enumerate}
\item The \textit{inferior cerebellar peduncle} (restiform body)
contains the dorsal spinocerebellar tract (DSCT) fibers. These
fibers arise from cells in the ipsilateral Clarke's column in the
spinal cord (C8--L3). This peduncle contains the cuneo--cerebellar
tract (CCT) fibers. These fibers arise from the ipsilateral
accessory cuneate nucleus. The largest component of the inferior
cerebellar peduncle consists of the olivo--cerebellar tract (OCT)
fibers. These fibers arise from the contralateral inferior olive.
Finally, vestibulo--cerebellar tract (VCT) fibers arise from cells
in both the vestibular ganglion and the vestibular nuclei and pass
in the inferior cerebellar peduncle to reach the cerebellum.

\item The \textit{middle cerebellar peduncle} (brachium pontis)
contains the pontocerebellar tract (PCT) fibers. These fibers
arise from the contralateral pontine grey.

\item The \textit{superior cerebellar peduncle} (brachium
conjunctivum) is the primary efferent (out of the cerebellum)
peduncle of the cerebellum. It contains fibers that arise from
several deep cerebellar nuclei. These fibers pass ipsilaterally
for a while and then cross at the level of the inferior colliculus
to form the decussation of the superior cerebellar peduncle. These
fibers then continue ipsilaterally to terminate in the red nucleus
(`ruber--duber') and the motor nuclei of the thalamus (VA, VL).
\end{enumerate}
\begin{figure}[ht]
\centerline{\includegraphics[width=8cm]{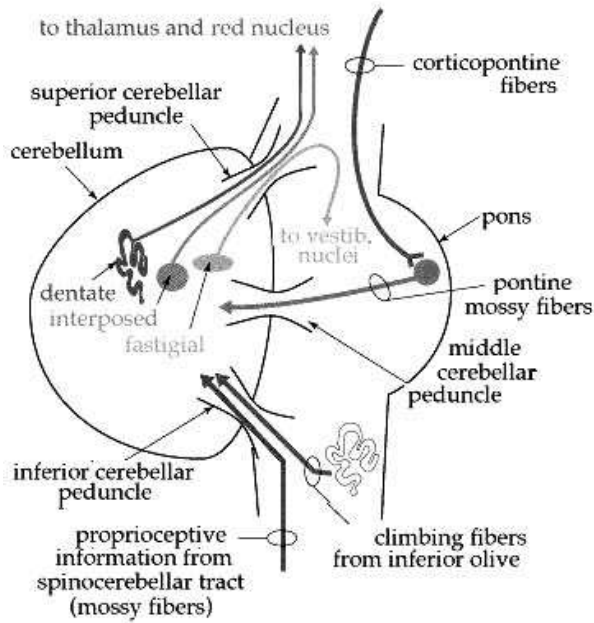}} \caption{Inputs and
outputs of the cerebellum.} \label{cb14}
\end{figure}

The 3 inputs are: \textit{mossy fibers} from the
\emph{spinocerebellar} pathways, climbing fibers from the
\emph{inferior olive}, and more mossy fibers from the \emph{pons},
which are carrying information from \emph{cerebral cortex} (see
Figure \ref{cb14}). The mossy fibers from the spinal cord have
come up ipsilaterally, so they do not need to cross. The fibers
coming down from cerebral cortex, however, do need to cross (as
the cerebrum is concerned with the opposite side of the body,
unlike the cerebellum). These fibers synapse in the pons (hence
the huge block of fibers in the cerebral peduncles labelled
`cortico--pontine'), cross, and enter the cerebellum as mossy
fibers.

The 3 deep nuclei are the \emph{fastigial}, \emph{interposed}, and
\emph{dentate nuclei}. The fastigial nucleus is primarily
concerned with balance, and sends information mainly to vestibular
and reticular nuclei. The dentate and interposed nuclei are
concerned more with voluntary movement, and send axons mainly to
thalamus and the red nucleus.

The main function of the cerebellum as a motor controller is
depicted in Figure \ref{CrblCtrl}. A coordinated movement is easy
to recognize, but we know little about how it is achieved. In
search of the neural basis of coordination, a model of
spinocerebellar interactions was recently presented in
\cite{Apps}, in which the structure–-functional organizing
principle is a division of the cerebellum into discrete
micro--complexes. Each micro--complex is the recipient of a
specific motor error signal, that is, a signal that conveys
information about an inappropriate movement. These signals are
encoded by spinal reflex circuits and conveyed to the cerebellar
cortex through climbing fibre afferents. This organization reveals
salient features of cerebellar information processing, but also
highlights the importance of systems level analysis for a fuller
understanding of the neural mechanisms that underlie behavior.

The authors of \cite{Apps} reviewed anatomical and physiological
foundations of cerebellar information processing. The cerebellum
is crucial for the coordination of movement. The authors presented
a model of the cerebellar paravermis, a region concerned with the
control of voluntary limb movements through its interconnections
with the spinal cord. They particularly focused on the
olivo–-cerebellar climbing fibre system.

Climbing fibres are proposed to convey motor error signals
(signals that convey information about inappropriate movements)
related to elementary limb movements that result from the
contraction of single muscles. The actual encoding of motor error
signals is suggested to depend on sensorimotor transformations
carried out by spinal modules that mediate nociceptive withdrawal
reflexes.

The termination of the climbing fibre system in the cerebellar
cortex subdivides the paravermis into distinct microzones.
Functionally similar but spatially separate microzones converge
onto a common group of cerebellar nuclear neurons. The processing
units formed as a consequence are termed `multizonal
micro-complexes' (MZMCs), and are each related to a specific
spinal reflex module.

The distributed nature of microzones that belong to a given MZMC
is proposed to enable similar climbing fibre inputs to integrate
with mossy fibre inputs that arise from different sources.
Anatomical results consistent with this notion have been obtained.

Within an individual MZMC, the skin receptive fields of climbing
fibres, mossy fibres and cerebellar cortical inhibitory
interneurons appear to be similar. This indicates that the
inhibitory receptive fields of Purkinje cells within a particular
MZMC result from the activation of inhibitory interneurons by
local granule cells.

On the other hand, the parallel fibre--mediated excitatory
receptive fields of the Purkinje cells in the same MZMC differ
from all of the other receptive fields, but are similar to those
of mossy fibres in another MZMC. This indicates that the
excitatory input to Purkinje cells in a given MZMC originates in
non--local granule cells and is mediated over some distance by
parallel fibres.

The output from individual MZMCs often involves two or three
segments of the ipsilateral limb, indicative of control of
multi--joint muscle synergies. The distal--most muscle in this
synergy seems to have a roughly antagonistic action to the muscle
associated with the climbing fibre input to the MZMC.

The model proposed in \cite{Apps} indicates that the cerebellar
paravermis system could provide the control of both single-- and
multi--joint movements. Agonist–-antagonist activity associated
with single--joint movements might be controlled within a
particular MZMC, whereas coordination across multiple joints might
be governed by interactions between MZMCs, mediated by parallel
fibres.

Two main theories address the function of the cerebellum, both
dealing with motor coordination. One claims that the cerebellum
functions as a regulator of the ``timing of movements." This has
emerged from studies of patients whose timed movements are
disrupted \cite{Ivry88}.

The second, ``Tensor Network Theory" provides a mathematical model
of transformation of sensory (covariant) space-time coordinates
into motor (contravariant) coordinates by cerebellar neuronal
networks \cite{Pellionisz80,Pellionisz82,Pellionisz85}.

Studies of motor learning in the vestibulo--ocular reflex and
eye-blink conditioning demonstrate that the timing and amplitude
of learned movements are encoded by the cerebellum
\cite{Boyden04}. Many synaptic plasticity mechanisms have been
found throughout the cerebellum. The \emph{Marr--Albus model}
mostly attributes motor learning to a single plasticity mechanism:
the long-term depression of parallel fiber synapses. The Tensor
Network Theory of sensory--motor transformations by the cerebellum
has also been experimentally supported \cite{Gielen86}.

\subsection{Feynman's Partition Function}

Recall that in statistical mechanics, the so--called \emph{partition function} $Z$ is a
quantity that encodes the statistical properties of a system in
thermodynamic equilibrium. It is a function of temperature and other
parameters, such as the volume enclosing a gas. Other thermodynamic
variables of the system, such as the total energy, free energy, entropy, and
pressure, can be expressed in terms of the partition function or its
derivatives.\footnote{%
There are actually several different types of partition functions, each
corresponding to different types of statistical ensemble (or, equivalently,
different types of free energy.) The canonical partition function applies to
a canonical ensemble, in which the system is allowed to exchange heat with
the environment at fixed temperature, volume, and number of particles. The
grand canonical partition function applies to a grand canonical ensemble, in
which the system can exchange both heat and particles with the environment,
at fixed temperature, volume, and chemical potential. Other types of
partition functions can be defined for different circumstances.}

The partition function of a \emph{canonical ensemble}\footnote{%
A canonical ensemble is a statistical ensemble representing a probability
distribution of microscopic states of the system. Its probability
distribution is characterized by the proportion $p_{i}$ of members of the
ensemble which exhibit a measurable macroscopic state $i$, where the
proportion of microscopic states for each macroscopic state $i$ is given by
the Boltzmann distribution,
\begin{equation*}
p_{i}=\tfrac{1}{Z}\mathrm{e}^{-E_{i}/(kT)}=\mathrm{e}^{-(E_{i}-A)/(kT)},
\end{equation*}%
where $E_{i}$ is the energy of state $i$. It can be shown that this is the
distribution which is most likely, if each system in the ensemble can
exchange energy with a heat bath, or alternatively with a large number of
similar systems. In other words, it is the distribution which has \emph{%
maximum entropy} for a given average energy $<E_{i}>$.} ~~is defined as a sum
~~$
Z(\beta )=\sum_{j}\mathrm{e}^{-\beta E_{j}},
$~
where $\beta =1/(k_{B}T)$ is the `inverse temperature', where $T$ is an
ordinary temperature and $k_{B}$ is the Boltzmann's constant. However, as
the position $x_{i}$ and momentum $p_{i}$ variables of an $i$th particle in
a system can vary continuously, the set of microstates is actually
uncountable. In this case, some form of \textit{coarse--graining} procedure
must be carried out, which essentially amounts to treating two mechanical
states as the same microstate if the differences in their position and
momentum variables are `small enough'. The partition function then takes the
form of an integral. For instance, the partition function of a gas
consisting of $N$ molecules is proportional to the $6N-$dimensional
phase--space integral,
\begin{equation*}
Z(\beta )\sim \int_{\mathbb{R}^{6N}}\,d^{3}p_{i}\,d^{3}x_{i}\exp [-\beta
H(p_{i},x_{i})],
\end{equation*}%
where $H=H(p_{i},x_{i}),$ ($i=1,...,N$) \ is the classical Hamiltonian
(total energy) function.

More generally, the so--called \textit{configuration integral}, as used in
probability theory, information science and dynamical systems, is an
abstraction of the above definition of a partition function in statistical
mechanics. It is a special case of a normalizing constant in probability
theory, for the Boltzmann distribution. The partition function occurs in
many problems of probability theory because, in situations where there is a
natural symmetry, its associated probability measure, the \emph{Gibbs measure%
} (see below), which generalizes the notion of the canonical ensemble, has
the \emph{Markov property}.

Given a set of random variables $X_{i}$ taking on values $x_{i}$, and purely
potential Hamiltonian function $H(x_{i}),$ ($i=1,...,N$), the partition
function is defined as
\begin{equation*}
Z(\beta )=\sum_{x_{i}}\exp \left[ -\beta H(x_{i})\right] .
\end{equation*}%
The function $H$ is understood to be a real-valued function on the space of
states $\{X_{1},X_{2},\cdots \}$ while $\beta $ is a real-valued free
parameter (conventionally, the inverse temperature). The sum over the $x_{i}$
is understood to be a sum over all possible values that the random variable $%
X_{i}$ may take. Thus, the sum is to be replaced by an integral when the $%
X_{i}$ are continuous, rather than discrete. Thus, one writes
\begin{equation*}
Z(\beta )=\int dx_{i}\exp \left[ -\beta H(x_{i})\right] ,
\end{equation*}%
for the case of continuously-varying random variables $X_{i}$.

The Gibbs measure of a random variable $X_{i}$ having the value $x_{i}$ is
defined as the probability density function
\begin{equation*}
P(X_{i}=x_{i})=\frac{1}{Z(\beta )}\exp \left[ -\beta E(x_{i})\right] =\frac{%
\exp \left[ -\beta H(x_{i})\right] }{\sum_{x_{i}}\exp \left[ -\beta H(x_{i})%
\right] }.
\end{equation*}%
where $E(x_{i})=H(x_{i})$ is the energy of the configuration $x_{i}$. This
probability, which is now properly normalized so that $0\leq P(x_{i})\leq 1,$
can be interpreted as a likelihood that a specific configuration of values $%
x_{i},$ ($i=1,2,...N$) occurs in the system.

As such, the partition function $Z(\beta )$\ can be understood to provide
the Gibbs measure on the space of states, which is the unique statistical
distribution that maximizes the entropy for a fixed expectation value of the
energy,
\begin{equation*}
\langle H\rangle =-\frac{\partial \log (Z(\beta ))}{\partial \beta }.
\end{equation*}%
The associated entropy is given by
\begin{equation*}
S=-\sum_{x_{i}}P(x_{i})\ln P(x_{i})=\beta \langle H\rangle +\log Z(\beta ).
\end{equation*}

The principle of maximum entropy related to the expectation value of the
energy $\langle H\rangle ,$ is a postulate about a universal feature of any
probability assignment on a given set of propositions (events, hypotheses,
indices, etc.). Let some testable information about a probability
distribution function be given. Consider the set of all trial probability
distributions which encode this information. Then the probability
distribution which maximizes the information entropy is the true probability
distribution, with respect to the testable information prescribed.

Now, the number of variables $X_{i}$ need not be countable, in which case
the set of coordinates $\{x_{i}\}$ becomes a field\ $\phi =\phi (x),$ so\
the sum is to be replaced by the \emph{Euclidean path integral} (that is a Wick--rotated
Feynman transition amplitude in imaginary time), as
\begin{equation*}
Z(\phi )=\int \mathcal{D}[\phi ]\exp \left[ -H(\phi )\right] .
\end{equation*}

More generally, in quantum field theory, instead of the field Hamiltonian $%
H(\phi )$ we have the action $S(\phi )$ of the theory. Both Euclidean
path integral,
\begin{equation}
Z(\phi )=\int \mathcal{D}[\phi ]\exp \left[ -S(\phi )\right] ,\qquad \text{%
real path integral in imaginary time}  \label{Eucl}
\end{equation}%
and Lorentzian one,
\begin{equation}
Z(\phi )=\int \mathcal{D}[\phi ]\exp \left[ iS(\phi )\right] ,\qquad \text{%
complex path integral in real time,}  \label{Lor}
\end{equation}%
are usually called `partition functions'. While the Lorentzian path integral (%
\ref{Lor}) represents a quantum-field theory-generalization of the Schr\"{o}%
dinger equation, the Euclidean path integral (\ref{Eucl}) represents a
statistical-field-theory generalization of the Fokker--Planck equation.


\begin{thebibliography}{Hat77a}
\bibitem[Mat69]{Matthews69} Matthews, P.B.C., Evidence that the secondary as
well as the primary endings of the muscle spindles may be responsible for
the tonic stretch-reflex of the decerebrate eat. J. Physiol. London \textbf{%
204}, 365--93, (1969)
\end{thebibliography}
\end{document}